\documentclass[conference,10pt,twoside,twocolumn]{IEEEtran}

\usepackage{amsmath,graphicx}

\usepackage[T1]{fontenc}
\usepackage{graphicx}
\usepackage{amssymb}
\usepackage{amsmath}
\usepackage{amsthm}
\usepackage{subfigure}
\usepackage{booktabs} 
\usepackage{multirow}
\usepackage{microtype}
\usepackage{cite}
\usepackage{subfigure}
\usepackage{xcolor} 
\usepackage{url}
\usepackage{balance}
\usepackage{colortbl}

\usepackage{amssymb}
\usepackage{amsfonts}
\usepackage{mathrsfs}
\usepackage{xspace}
\usepackage{bm}
\usepackage{upgreek}

\newcommand{\safemath}[2]{\newcommand{#1}{\ensuremath{#2}\xspace}}

\safemath{\bma}{\mathbf{a}}
\safemath{\bmb}{\mathbf{b}}
\safemath{\bmc}{\mathbf{c}}
\safemath{\bmd}{\mathbf{d}}
\safemath{\bme}{\mathbf{e}}
\safemath{\bmf}{\mathbf{f}}
\safemath{\bmg}{\mathbf{g}}
\safemath{\bmh}{\mathbf{h}}
\safemath{\bmi}{\mathbf{i}}
\safemath{\bmj}{\mathbf{j}}
\safemath{\bmk}{\mathbf{k}}
\safemath{\bml}{\mathbf{l}}
\safemath{\bmm}{\mathbf{m}}
\safemath{\bmn}{\mathbf{n}}
\safemath{\bmo}{\mathbf{o}}
\safemath{\bmp}{\mathbf{p}}
\safemath{\bmq}{\mathbf{q}}
\safemath{\bmr}{\mathbf{r}}
\safemath{\bms}{\mathbf{s}}
\safemath{\bmt}{\mathbf{t}}
\safemath{\bmu}{\mathbf{u}}
\safemath{\bmv}{\mathbf{v}}
\safemath{\bmw}{\mathbf{w}}
\safemath{\bmx}{\mathbf{x}}
\safemath{\bmy}{\mathbf{y}}
\safemath{\bmz}{\mathbf{z}}
\safemath{\bmzero}{\mathbf{0}}
\safemath{\bmone}{\mathbf{1}}
\safemath{\Bell}{\ensuremath{\boldsymbol\ell}}

\bmdefine{\biad}{a}
\bmdefine{\bibd}{b}
\bmdefine{\bicd}{c}
\bmdefine{\bidd}{d}
\bmdefine{\bied}{e}
\bmdefine{\bifd}{f}
\bmdefine{\bigd}{g}
\bmdefine{\bihd}{h}
\bmdefine{\biid}{i}
\bmdefine{\bijd}{j}
\bmdefine{\bikd}{k}
\bmdefine{\bild}{l}
\bmdefine{\bimd}{m}
\bmdefine{\bind}{n}
\bmdefine{\biod}{o}
\bmdefine{\bipd}{p}
\bmdefine{\biqd}{q}
\bmdefine{\bird}{r}
\bmdefine{\bisd}{s}
\bmdefine{\bitd}{t}
\bmdefine{\biud}{u}
\bmdefine{\bivd}{v}
\bmdefine{\biwd}{w}
\bmdefine{\bixd}{x}
\bmdefine{\biyd}{y}
\bmdefine{\bizd}{z}

\bmdefine{\bixid}{\xi}
\bmdefine{\bilambdad}{\lambda}
\bmdefine{\bimud}{\mu}
\bmdefine{\bithetad}{\theta}
\bmdefine{\biphid}{\phi}
\bmdefine{\bideltad}{\delta}

\safemath{\bmia}{\biad}
\safemath{\bmib}{\bibd}
\safemath{\bmic}{\bicd}
\safemath{\bmid}{\bidd}
\safemath{\bmie}{\bied}
\safemath{\bmif}{\bifd}
\safemath{\bmig}{\bigd}
\safemath{\bmih}{\bihd}
\safemath{\bmii}{\biid}
\safemath{\bmij}{\bijd}
\safemath{\bmik}{\bikd}
\safemath{\bmil}{\bild}
\safemath{\bmim}{\bimd}
\safemath{\bmin}{\bind}
\safemath{\bmio}{\biod}
\safemath{\bmip}{\bipd}
\safemath{\bmiq}{\biqd}
\safemath{\bmir}{\bird}
\safemath{\bmis}{\bisd}
\safemath{\bmit}{\bitd}
\safemath{\bmiu}{\biud}
\safemath{\bmiv}{\bivd}
\safemath{\bmiw}{\biwd}
\safemath{\bmix}{\bixd}
\safemath{\bmiy}{\biyd}
\safemath{\bmiz}{\bizd}

\safemath{\bmxi}{\bixid}
\safemath{\bmlambda}{\bilambdad}
\safemath{\bmmu}{\bimud}
\safemath{\bmtheta}{\bithetad}
\safemath{\bmphi}{\biphid}
\safemath{\bmdelta}{\bideltad}

\safemath{\bA}{\mathbf{A}}
\safemath{\bB}{\mathbf{B}}
\safemath{\bC}{\mathbf{C}}
\safemath{\bD}{\mathbf{D}}
\safemath{\bE}{\mathbf{E}}
\safemath{\bF}{\mathbf{F}}
\safemath{\bG}{\mathbf{G}}
\safemath{\bH}{\mathbf{H}}
\safemath{\bI}{\mathbf{I}}
\safemath{\bJ}{\mathbf{J}}
\safemath{\bK}{\mathbf{K}}
\safemath{\bL}{\mathbf{L}}
\safemath{\bM}{\mathbf{M}}
\safemath{\bN}{\mathbf{N}}
\safemath{\bO}{\mathbf{O}}
\safemath{\bP}{\mathbf{P}}
\safemath{\bQ}{\mathbf{Q}}
\safemath{\bR}{\mathbf{R}}
\safemath{\bS}{\mathbf{S}}
\safemath{\bT}{\mathbf{T}}
\safemath{\bU}{\mathbf{U}}
\safemath{\bV}{\mathbf{V}}
\safemath{\bW}{\mathbf{W}}
\safemath{\bX}{\mathbf{X}}
\safemath{\bY}{\mathbf{Y}}
\safemath{\bZ}{\mathbf{Z}}

\safemath{\bZero}{\mathbf{0}}
\safemath{\bOne}{\mathbf{1}}
\safemath{\bDelta}{\mathbf{\Delta}}
\safemath{\bLambda}{\mathbf{\UpLambda}}
\safemath{\bPhi}{\mathbf{\Upphi}}
\safemath{\bSigma}{\mathbf{\Upsigma}}
\safemath{\bOmega}{\mathbf{\Upomega}}
\safemath{\bTheta}{\mathbf{\Uptheta}}

\bmdefine{\biAd}{A}
\bmdefine{\biBd}{B}
\bmdefine{\biCd}{C}
\bmdefine{\biDd}{D}
\bmdefine{\biEd}{E}
\bmdefine{\biFd}{F}
\bmdefine{\biGd}{G}
\bmdefine{\biHd}{H}
\bmdefine{\biId}{I}
\bmdefine{\biJd}{J}
\bmdefine{\biKd}{K}
\bmdefine{\biLd}{L}
\bmdefine{\biMd}{M}
\bmdefine{\biOd}{N}
\bmdefine{\biPd}{O}
\bmdefine{\biQd}{P}
\bmdefine{\biRd}{R}
\bmdefine{\biSd}{S}
\bmdefine{\biTd}{T}
\bmdefine{\biUd}{U}
\bmdefine{\biVd}{V}
\bmdefine{\biWd}{W}
\bmdefine{\biXd}{X}
\bmdefine{\biYd}{Y}
\bmdefine{\biZd}{Z}

\bmdefine{\biDelta}{\Delta}
\bmdefine{\biLambda}{\Lambda}
\bmdefine{\biPhi}{\Phi}
\bmdefine{\biSigma}{\Sigma}
\bmdefine{\biOmega}{\Omega}
\bmdefine{\biTheta}{\Theta}

\safemath{\bimA}{\biAd}
\safemath{\bimB}{\biBd}
\safemath{\bimC}{\biCd}
\safemath{\bimD}{\biDd}
\safemath{\bimE}{\biEd}
\safemath{\bimF}{\biFd}
\safemath{\bimG}{\biGd}
\safemath{\bimH}{\biHd}
\safemath{\bimI}{\biId}
\safemath{\bimJ}{\biJd}
\safemath{\bimK}{\biKd}
\safemath{\bimL}{\biLd}
\safemath{\bimM}{\biMd}
\safemath{\bimN}{\biNd}
\safemath{\bimO}{\biOd}
\safemath{\bimP}{\biPd}
\safemath{\bimQ}{\biQd}
\safemath{\bimR}{\biRd}
\safemath{\bimS}{\biSd}
\safemath{\bimT}{\biTd}
\safemath{\bimU}{\biUd}
\safemath{\bimV}{\biVd}
\safemath{\bimW}{\biWd}
\safemath{\bimX}{\biXd}
\safemath{\bimY}{\biYd}
\safemath{\bimZ}{\biZd}

\safemath{\bimDelta}{\biDelta}
\safemath{\bimLambda}{\biLambda}
\safemath{\bimPhi}{\biPhi}
\safemath{\bimSigma}{\biSigma}
\safemath{\bimOmega}{\biOmega}
\safemath{\bimTheta}{\biTheta}

\safemath{\setA}{\mathcal{A}}
\safemath{\setB}{\mathcal{B}}
\safemath{\setC}{\mathcal{C}}
\safemath{\setD}{\mathcal{D}}
\safemath{\setE}{\mathcal{E}}
\safemath{\setF}{\mathcal{F}}
\safemath{\setG}{\mathcal{G}}
\safemath{\setH}{\mathcal{H}}
\safemath{\setI}{\mathcal{I}}
\safemath{\setJ}{\mathcal{J}}
\safemath{\setK}{\mathcal{K}}
\safemath{\setL}{\mathcal{L}}
\safemath{\setM}{\mathcal{M}}
\safemath{\setN}{\mathcal{N}}
\safemath{\setO}{\mathcal{O}}
\safemath{\setP}{\mathcal{P}}
\safemath{\setQ}{\mathcal{Q}}
\safemath{\setR}{\mathcal{R}}
\safemath{\setS}{\mathcal{S}}
\safemath{\setT}{\mathcal{T}}
\safemath{\setU}{\mathcal{U}}
\safemath{\setV}{\mathcal{V}}
\safemath{\setW}{\mathcal{W}}
\safemath{\setX}{\mathcal{X}}
\safemath{\setY}{\mathcal{Y}}
\safemath{\setZ}{\mathcal{Z}}
\safemath{\emptySet}{\varnothing}

\safemath{\colA}{\mathscr{A}}
\safemath{\colB}{\mathscr{B}}
\safemath{\colC}{\mathscr{C}}
\safemath{\colD}{\mathscr{D}}
\safemath{\colE}{\mathscr{E}}
\safemath{\colF}{\mathscr{F}}
\safemath{\colG}{\mathscr{G}}
\safemath{\colH}{\mathscr{H}}
\safemath{\colI}{\mathscr{I}}
\safemath{\colJ}{\mathscr{J}}
\safemath{\colK}{\mathscr{K}}
\safemath{\colL}{\mathscr{L}}
\safemath{\colM}{\mathscr{M}}
\safemath{\colN}{\mathscr{N}}
\safemath{\colO}{\mathscr{O}}
\safemath{\colP}{\mathscr{P}}
\safemath{\colQ}{\mathscr{Q}}
\safemath{\colR}{\mathscr{R}}
\safemath{\colS}{\mathscr{S}}
\safemath{\colT}{\mathscr{T}}
\safemath{\colU}{\mathscr{U}}
\safemath{\colV}{\mathscr{V}}
\safemath{\colW}{\mathscr{W}}
\safemath{\colX}{\mathscr{X}}
\safemath{\colY}{\mathscr{Y}}
\safemath{\colZ}{\mathscr{Z}}

\safemath{\opA}{\mathbb{A}}
\safemath{\opB}{\mathbb{B}}
\safemath{\opC}{\mathbb{C}}
\safemath{\opD}{\mathbb{D}}
\safemath{\opE}{\mathbb{E}}
\safemath{\opF}{\mathbb{F}}
\safemath{\opG}{\mathbb{G}}
\safemath{\opH}{\mathbb{H}}
\safemath{\opI}{\mathbb{I}}
\safemath{\opJ}{\mathbb{J}}
\safemath{\opK}{\mathbb{K}}
\safemath{\opL}{\mathbb{L}}
\safemath{\opM}{\mathbb{M}}
\safemath{\opN}{\mathbb{N}}
\safemath{\opO}{\mathbb{O}}
\safemath{\opP}{\mathbb{P}}
\safemath{\opQ}{\mathbb{Q}}
\safemath{\opR}{\mathbb{R}}
\safemath{\opS}{\mathbb{S}}
\safemath{\opT}{\mathbb{T}}
\safemath{\opU}{\mathbb{U}}
\safemath{\opV}{\mathbb{V}}
\safemath{\opW}{\mathbb{W}}
\safemath{\opX}{\mathbb{X}}
\safemath{\opY}{\mathbb{Y}}
\safemath{\opZ}{\mathbb{Z}}
\safemath{\opZero}{\mathbb{O}}
\safemath{\identityop}{\opI}

\safemath{\veca}{\bma}
\safemath{\vecb}{\bmb}
\safemath{\vecc}{\bmc}
\safemath{\vecd}{\bmd}
\safemath{\vece}{\bme}
\safemath{\vecf}{\bmf}
\safemath{\vecg}{\bmg}
\safemath{\vech}{\bmh}
\safemath{\veci}{\bmi}
\safemath{\vecj}{\bmj}
\safemath{\veck}{\bmk}
\safemath{\vecl}{\bml}
\safemath{\vecm}{\bmm}
\safemath{\vecn}{\bmn}
\safemath{\veco}{\bmo}
\safemath{\vecp}{\bmp}
\safemath{\vecq}{\bmq}
\safemath{\vecr}{\bmr}
\safemath{\vecs}{\bms}
\safemath{\vect}{\bmt}
\safemath{\vecu}{\bmu}
\safemath{\vecv}{\bmv}
\safemath{\vecw}{\bmw}
\safemath{\vecx}{\bmx}
\safemath{\vecy}{\bmy}
\safemath{\vecz}{\bmz}

\safemath{\veczero}{\bmzero}
\safemath{\vecone}{\bmone}
\safemath{\vecxi}{\bmxi}
\safemath{\veclambda}{\bmlambda}
\safemath{\vecmu}{\bmmu}
\safemath{\vectheta}{\bmtheta}
\safemath{\vecphi}{\bmphi}
\safemath{\vecdelta}{\bmdelta}

\safemath{\matA}{\bA}
\safemath{\matB}{\bB}
\safemath{\matC}{\bC}
\safemath{\matD}{\bD}
\safemath{\matE}{\bE}
\safemath{\matF}{\bF}
\safemath{\matG}{\bG}
\safemath{\matH}{\bH}
\safemath{\matI}{\bI}
\safemath{\matJ}{\bJ}
\safemath{\matK}{\bK}
\safemath{\matL}{\bL}
\safemath{\matM}{\bM}
\safemath{\matN}{\bN}
\safemath{\matO}{\bO}
\safemath{\matP}{\bP}
\safemath{\matQ}{\bQ}
\safemath{\matR}{\bR}
\safemath{\matS}{\bS}
\safemath{\matT}{\bT}
\safemath{\matU}{\bU}
\safemath{\matV}{\bV}
\safemath{\matW}{\bW}
\safemath{\matX}{\bX}
\safemath{\matY}{\bY}
\safemath{\matZ}{\bZ}
\safemath{\matzero}{\bmzero}

\safemath{\matDelta}{\bDelta}
\safemath{\matLambda}{\bLambda}
\safemath{\matPhi}{\bPhi}
\safemath{\matSigma}{\bSigma}
\safemath{\matOmega}{\bOmega}
\safemath{\matTheta}{\bTheta}

\safemath{\matidentity}{\matI}
\safemath{\matone}{\matO}

\safemath{\rnda}{A}
\safemath{\rndb}{B}
\safemath{\rndc}{C}
\safemath{\rndd}{D}
\safemath{\rnde}{E}
\safemath{\rndf}{F}
\safemath{\rndg}{G}
\safemath{\rndh}{H}
\safemath{\rndi}{I}
\safemath{\rndj}{J}
\safemath{\rndk}{K}
\safemath{\rndl}{L}
\safemath{\rndm}{M}
\safemath{\rndn}{N}
\safemath{\rndo}{O}
\safemath{\rndp}{P}
\safemath{\rndq}{Q}
\safemath{\rndr}{R}
\safemath{\rnds}{S}
\safemath{\rndt}{T}
\safemath{\rndu}{U}
\safemath{\rndv}{V}
\safemath{\rndw}{W}
\safemath{\rndx}{X}
\safemath{\rndy}{Y}
\safemath{\rndz}{Z}

\safemath{\rveca}{\bimA}
\safemath{\rvecb}{\bimB}
\safemath{\rvecc}{\bimC}
\safemath{\rvecd}{\bimD}
\safemath{\rvece}{\bimE}
\safemath{\rvecf}{\bimF}
\safemath{\rvecg}{\bimG}
\safemath{\rvech}{\bimH}
\safemath{\rveci}{\bimI}
\safemath{\rvecj}{\bimJ}
\safemath{\rveck}{\bimK}
\safemath{\rvecl}{\bimL}
\safemath{\rvecm}{\bimM}
\safemath{\rvecn}{\bimN}
\safemath{\rveco}{\bomO}
\safemath{\rvecp}{\bimP}
\safemath{\rvecq}{\bimQ}
\safemath{\rvecr}{\bimR}
\safemath{\rvecs}{\bimS}
\safemath{\rvect}{\bimT}
\safemath{\rvecu}{\bimU}
\safemath{\rvecv}{\bimV}
\safemath{\rvecw}{\bimW}
\safemath{\rvecx}{\bimX}
\safemath{\rvecy}{\bimY}
\safemath{\rvecz}{\bimZ}

\safemath{\rvecxi}{\bmxi}
\safemath{\rveclambda}{\bmlambda}
\safemath{\rvecmu}{\bmmu}
\safemath{\rvectheta}{\bmtheta}
\safemath{\rvecphi}{\bmphi}

\safemath{\rmatA}{\bimA}
\safemath{\rmatB}{\bimB}
\safemath{\rmatC}{\bimC}
\safemath{\rmatD}{\bimD}
\safemath{\rmatE}{\bimE}
\safemath{\rmatF}{\bimF}
\safemath{\rmatG}{\bimG}
\safemath{\rmatH}{\bimH}
\safemath{\rmatI}{\bimI}
\safemath{\rmatJ}{\bimJ}
\safemath{\rmatK}{\bimK}
\safemath{\rmatL}{\bimL}
\safemath{\rmatM}{\bimM}
\safemath{\rmatN}{\bimN}
\safemath{\rmatO}{\bimO}
\safemath{\rmatP}{\bimP}
\safemath{\rmatQ}{\bimQ}
\safemath{\rmatR}{\bimR}
\safemath{\rmatS}{\bimS}
\safemath{\rmatT}{\bimT}
\safemath{\rmatU}{\bimU}
\safemath{\rmatV}{\bimV}
\safemath{\rmatW}{\bimW}
\safemath{\rmatX}{\bimX}
\safemath{\rmatY}{\bimY}
\safemath{\rmatZ}{\bimZ}

\safemath{\rmatDelta}{\bimDelta}
\safemath{\rmatLambda}{\bimLambda}
\safemath{\rmatPhi}{\bimPhi}
\safemath{\rmatSigma}{\bimSigma}
\safemath{\rmatOmega}{\bimOmega}
\safemath{\rmatTheta}{\bimTheta}

\usepackage{amssymb}
\usepackage{amsfonts}
\usepackage{mathrsfs}
\usepackage{xspace}
\usepackage{bm}
\usepackage{fancyref}
\usepackage{textcomp}

\usepackage{multirow}
\usepackage{stmaryrd}

\newenvironment{textbmatrix}{	\setlength{\arraycolsep}{2.5pt}%
								\left[\begin{matrix}}{\end{matrix}\right]%
								\raisebox{0.08ex}{\vphantom{M}}}

\def\be{\begin{equation}}
\def\ee{\end{equation}}
\def\een{\nonumber \end{equation}}
\def\mat{\begin{bmatrix}}
\def\emat{\end{bmatrix}}
\def\btm{\begin{textbmatrix}}
\def\etm{\end{textbmatrix}}

\def\ba#1\ea{\begin{align}#1\end{align}}
\def\bas#1\eas{\begin{align*}#1\end{align*}}
\def\bs#1\es{\begin{split}#1\end{split}}
\def\bg#1\eg{\begin{gather}#1\end{gather}}
\def\bml#1\eml{\begin{multline}#1\end{multline}}
\def\bi#1\ei{\begin{itemize}#1\end{itemize}}

\newcommand{\lefto}{\mathopen{}\left}

\DeclareMathOperator{\rect}{rect}			
\DeclareMathOperator{\Exop}{\opE}			

\newcommand{\Ex}[1]{\ensuremath{\Exop\lefto[#1\right]}} 	

\newcommand{\vecnorm}[1]{\lefto\lVert#1\right\rVert}		
\newcommand{\frobnorm}[1]{\vecnorm{#1}_{\text{F}}}	

\safemath{\dirac}{\delta}					
\safemath{\krond}{\dirac}					

\safemath{\upto}{\uparrow}
\safemath{\downto}{\downarrow}
\safemath{\iu}{j}							
\safemath{\ev}{\lambda}						
\safemath{\hilseqspace}{l^{2}}				
\newcommand{\banachfunspace}[1]{\setL^{#1}}	
\safemath{\hilfunspace}{\banachfunspace{2}}	

\safemath{\SNR}{\textit{SNR}} 				
\safemath{\PAR}{\textit{PAR}} 				
\safemath{\No}{N_0}							
\safemath{\Es}{E_s}							
\safemath{\Eb}{E_b}							
\safemath{\EbNo}{\frac{\Eb}{\No}}
\safemath{\EsNo}{\frac{\Es}{\No}}

\DeclareMathOperator{\CHop}{\ensuremath{\opH}} 
\safemath{\tvir}{\rndh_{\CHop}}				
\safemath{\tvtf}{\rndl_{\CHop}}				
\safemath{\spf}{\rnds_{\CHop}}				
\safemath{\bff}{H_{\CHop}}					

\safemath{\ircf}{r_{h}}						
\safemath{\tftvcf}{r_{s}}					
\safemath{\tfcf}{r_{l}}						
\safemath{\bfcf}{r_{H}}						

\safemath{\tcorr}{c_h}						
\safemath{\scf}{c_{s}}						
\safemath{\tfcorr}{c_{l}}					
\safemath{\fcorr}{c_{H}}						

\safemath{\mi}{I}							
\safemath{\capacity}{C}						

\safemath{\normal}{\mathcal{N}}			
\safemath{\jpg}{\mathcal{CN}}			
\safemath{\mchain}{\leftrightarrow}		

\safemath{\dB}{\,\mathrm{dB}}
\safemath{\dBm}{\,\mathrm{dBm}}
\safemath{\Hz}{\,\mathrm{Hz}}
\safemath{\kHz}{\,\mathrm{kHz}}
\safemath{\MHz}{\,\mathrm{MHz}}
\safemath{\GHz}{\,\mathrm{GHz}}
\safemath{\s}{\,\mathrm{s}}
\safemath{\ms}{\,\mathrm{ms}}
\safemath{\mus}{\,\mathrm{\text{\textmu}s}}
\safemath{\ns}{\,\mathrm{ns}}
\safemath{\ps}{\,\mathrm{ps}}
\safemath{\meter}{\,\mathrm{m}}
\safemath{\mm}{\,\mathrm{mm}}
\safemath{\cm}{\,\mathrm{cm}}
\safemath{\m}{\,\mathrm{m}}
\safemath{\W}{\,\mathrm{W}}
\safemath{\mW}{\, \mathrm{mW}}
\safemath{\J}{\,\mathrm{J}}
\safemath{\K}{\,\mathrm{K}}
\safemath{\bit}{\,\mathrm{bit}}
\safemath{\nat}{\,\mathrm{nat}}

\safemath{\define}{\triangleq}			

\safemath{\equivalent}{\sim}
\safemath{\distas}{\sim}					
\safemath{\sdiff}{\Delta}				

\safemath{\reals}{\mathbb{R}}
\safemath{\positivereals}{\reals_{+}}
\safemath{\integers}{\mathbb{Z}}
\safemath{\posint}{\integers_{+}}
\safemath{\naturals}{\mathbb{N}}
\safemath{\posnaturals}{\naturals_{+}}
\safemath{\complexset}{\mathbb{C}}
\safemath{\rationals}{\mathbb{Q}}

\newcommand*{\fancyrefapplabelprefix}{app}		
\newcommand*{\fancyrefthmlabelprefix}{thm}		
\newcommand*{\fancyreflemlabelprefix}{lem}		
\newcommand*{\fancyrefcorlabelprefix}{cor}		
\newcommand*{\fancyrefdeflabelprefix}{def}		
\newcommand*{\fancyrefproplabelprefix}{prop}		
\newcommand*{\fancyrefexmpllabelprefix}{exmpl}
\newcommand*{\fancyrefalglabelprefix}{alg}		
\newcommand*{\fancyreftbllabelprefix}{tbl}		

\frefformat{vario}{\fancyrefseclabelprefix}{Sec.~#1}
\frefformat{vario}{\fancyrefthmlabelprefix}{Theorem~#1}
\frefformat{vario}{\fancyreftbllabelprefix}{Tbl.~#1}
\frefformat{vario}{\fancyreflemlabelprefix}{Lemma~#1}
\frefformat{vario}{\fancyrefcorlabelprefix}{Corollary~#1}
\frefformat{vario}{\fancyrefdeflabelprefix}{Definition~#1}
\frefformat{vario}{\fancyreffiglabelprefix}{Fig.~#1}
\frefformat{vario}{\fancyrefapplabelprefix}{Appendix~#1}
\frefformat{vario}{\fancyrefeqlabelprefix}{(#1)}
\frefformat{vario}{\fancyrefproplabelprefix}{Proposition~#1}
\frefformat{vario}{\fancyrefexmpllabelprefix}{Example~#1}
\frefformat{vario}{\fancyrefalglabelprefix}{Algorithm~#1}

\safemath{\dictab}{[\,\dicta\,\,\dictb\,]}

\safemath{\ysig}{\bmy}
\safemath{\ysighat}{\hat{\ysig}}
\safemath{\ysigdim}{M}
\safemath{\xsig}{\bmx}
\safemath{\xsigdim}{N}
\safemath{\nx}{n_x}
\safemath{\zsig}{\bmz}
\safemath{\zsigdim}{\ysigdim}
\safemath{\rsig}{\bmr}
\safemath{\Adict}{\bA}
\safemath{\Adicttilde}{\widetilde{\Adict}}
\safemath{\Adictdim}{\outputdim\times\xsigdim}
\safemath{\avec}{\bma}
\safemath{\avectilde}{\tilde{\avec}}
\safemath{\Bdict}{\bB}
\safemath{\Bdicttilde}{\widetilde{\Bdict}}
\safemath{\Cdict}{\bC}
\safemath{\cvec}{\bmc}
\safemath{\Ddict}{\bD}
\safemath{\Ddictdim}{\ysigdim\times\xsigdim}
\safemath{\dvec}{\bmd}
\safemath{\Ddicttilde}{\widetilde{\bD}}
\safemath{\Bonb}{\bB}
\safemath{\bvec}{\bmb}
\safemath{\Bonbdim}{\ysigdim\times\ysigdim}
\safemath{\noise}{\bmn}
\safemath{\noisedim}{\ysigim}
\safemath{\err}{\bme}
\safemath{\errdim}{\ysigdim}
\safemath{\errset}{\setE}
\safemath{\nerr}{n_e}
\safemath{\delop}{\bP_\errset}
\safemath{\delopc}{\bP_{{\errset}^c}}

\safemath{\cplxi}{\imath}
\safemath{\cplxj}{\jmath}

\safemath{\dict}{\matD}
\safemath{\inputdim}{N}		
\safemath{\outputdim}{M}		
\safemath{\sparsity}{S}	
\safemath{\inputdimA}{{N_a}}	
\safemath{\inputdimB}{{N_b}}	
\safemath{\elemA}{{n_a}}	
\safemath{\elemB}{{n_b}}	
\safemath{\resA}{\matR_a}	
\safemath{\resB}{\matR_b}	
\safemath{\subD}{\matS} 
\safemath{\subA}{\matS_a} 
\safemath{\subB}{\matS_b} 
\safemath{\dicta}{\matA} 	
\safemath{\dictb}{\matB} 	
\safemath{\hollowS}{H}
\safemath{\hollowA}{H_a}
\safemath{\hollowB}{H_b}
\safemath{\cross}{Z}
\safemath{\coh}{\mu_d}			
\safemath{\coha}{\mu_a}			
\safemath{\cohb}{\mu_b}			
\safemath{\mubs}{\nu}	
\safemath{\cohm}{\mu_m} 
\safemath{\dictset}{\setD}	
\safemath{\dictsetp}{\dictset(\coh,\coha,\cohb)}	
\safemath{\dictsetgen}{\dictset_\text{gen}}
\safemath{\dictsetgenp}{\dictsetgen(\coh)}
\safemath{\dictsetonb}{\dictset_\text{onb}}
\safemath{\dictsetonbp}{\dictsetonb(\coh)}

\safemath{\leftside}{U}
\safemath{\rightsideA}{R_a}
\safemath{\rightsideB}{R_b}

\safemath{\indexS}{\setI_S} 

\safemath{\na}{n_a}			
\safemath{\nb}{n_b}			
\safemath{\coeffa}{p_i}	
\safemath{\coeffb}{q_j}	
\safemath{\seta}{\setP}		
\safemath{\setb}{\setQ}     
\safemath{\setw}{\setW}	
\safemath{\setz}{\setZ}	
\safemath{\cola}{\veca}		
\safemath{\colb}{\vecb}		
\safemath{\cold}{\vecd}		
\safemath{\inputvec}{\vecx} 	
\safemath{\error}{\vece}	
\safemath{\noiseout}{\vecz} 	
\safemath{\inputvecel}{x}
\safemath{\inputveca}{\vecx_a}
\safemath{\inputvecb}{\vecx_b}
\safemath{\outputvec}{\vecy}	
\safemath{\lambdamin}{\lambda_{\mathrm{min}}}

\safemath{\elltwo}{\ell_2}
\safemath{\ellone}{\ell_1}
\safemath{\ellzero}{\ell_0}
\safemath{\ellinf}{\ell_\infty}
\safemath{\ellinftilde}{\ell_{\widetilde\infty}}
\safemath{\licard}{Z(\coh,\coha,\cohb)}
\safemath{\xsol}{\hat{x}}
\safemath{\xbord}{x_b}		
\safemath{\xstat}{x_s}		
\safemath{\xstatLone}{\tilde{x}_s}
\safemath{\order}{\mathcal{O}} 
\safemath{\scales}{\Theta} 
\safemath{\ones}{\mathbf{1}} 
\safemath{\zeroes}{\mathbf{0}} 
\safemath{\thlone}{\kappa(\coh,\cohb)} 
\safemath{\constoneA}{\delta} 
\safemath{\constoneB}{\epsilon} 
\safemath{\nlarge}{L}				   
\safemath{\sumlarge}{S_\nlarge}
\safemath{\maxlarger}{P_\nlarge}	   
\safemath{\Pzero}{\textrm{P0}}	
\safemath{\Pone}{\textrm{P1}}
\safemath{\vecfir}{\vecw}			 
\safemath{\vecsec}{\vecz}
\safemath{\elvecfir}{w}              
\safemath{\elvecsec}{z}				 
\safemath{\nlargefir}{n}
\safemath{\normout}{\gamma}
\safemath{\auxfun}{h}
\safemath{\supp}{\textrm{supp}}

\safemath{\indexa}{\ell}
\safemath{\indexb}{r}
\safemath{\indexc}{i}
\safemath{\indexd}{j}

\safemath{\project}{P}

\usepackage{framed}

\IEEEoverridecommandlockouts
\allowdisplaybreaks 

\begin{document}

\title{The Impact of SAR-ADC Mismatch on \\ Quantized Massive MU-MIMO Systems}

\author{J\'er\'emy Guichemerre and Christoph Studer\\[0.3cm]
\textit{Dept. of Information Technology and Electrical Engineering, ETH Zurich, Switzerland} \\ 
\textit{email: jeremyg@iis.ee.ethz.ch and studer@ethz.ch}
\thanks{The work of JG and CS was supported in part by an ETH Research Grant. The work of CS was supported in part by the U.S. National Science Foundation (NSF) under grants CNS-1717559 and ECCS-1824379. }
\thanks{The authors would like to thank Dr.~Thomas Burger for his support, Gian Marti for comments on an initial version of the paper, as well as Seyed Hadi Mirfarshbafan and Darja Nonaca for their help with the system simulator.}
}

\maketitle


\begin{abstract}
Low-resolution analog-to-digital converters (ADCs) in massive multi-user (MU) multiple-input multiple-output (MIMO) wireless systems can significantly reduce the power, cost, and interconnect data rates  of infrastructure basestations. Thus, recent research on the theory and algorithm sides has extensively focused on such architectures, but with idealistic quantization models. However, real-world ADCs do \emph{not} behave like ideal quantizers, and are affected by fabrication mismatches. We analyze the impact of capacitor-array mismatches in successive approximation register (SAR) ADCs, which are widely used in wireless systems. We use Bussgang's decomposition to model the effects of such mismatches, and we analyze their impact on the performance of a single ADC. We then simulate a  massive MU-MIMO system to demonstrate that capacitor mismatches should \emph{not} be ignored, even in basestations that use low-resolution SAR ADCs.
\end{abstract}


\section{Introduction}

Millimeter-wave frequencies provide access to large portions of bandwidth and are therefore targeted for fifth-generation (5G) and beyond-5G wireless systems\mbox{\cite{Rappaport13,Swindlehurst14}}.
Combined with multi-user (MU) massive multiple-input multiple-output (MIMO), one cannot only achieve high-data-rate communication with  multiple user equipments (UEs) in the same frequency band, but also combat the strong path loss at such carrier frequencies by means of beamforming~\cite{Lu14,Bjornsso17}.
However, the large number of antenna elements and radio-frequency chains in massive MU-MIMO basestation architectures combined with large bandwidths poses significant practical implementation challenges in terms of system costs and power consumption. 

Quantized massive MU-MIMO~\cite{Mollen17,Jacobsson17,Abdallah18} has become a popular means to reduce the costs and power consumption in the analog front-ends (FEs).
In fact, decreasing the analog-to-digital converter (ADC) resolution relaxes the design requirements of the entire analog FE, thereby reducing power consumption and silicon area for both analog FEs and digital processing.
Thus, research has mainly focused on understanding the effects of coarse quantization and on the design of corresponding baseband algorithms that mitigate quantization artifacts.
However, the impact of quantization non-idealities that arise in real-world ADCs has been routinely ignored in the literature.

The successive approximation register (SAR) ADC~\cite{Suarez74} is the data converter of choice for many communications applications supporting bandwidths of tens of megahertz to low gigahertz\mbox{\cite{adc_survey,Walden99}}.
The popularity of SAR ADCs is due to several reasons: First, they require low energy per conversion step. Second, they scale well to smaller semiconductor technology nodes. Third, they are suitable for pipelined and interleaved architectures, which enable one to achieve even higher sampling rates~\cite{Jiang20,Pfaff19,Lagos22}.
However, as all real-world data converters, SAR ADCs are \emph{not} ideal quantizers and are affected by imperfections, which already occur during manufacturing.
Specifically, the capacitors used for data conversion suffer from mismatches, i.e., the capacitor values of the fabricated circuit deviate from the specified ideal ones, which results in distortions of the quantization function. 
The impact of such capacitor mismatches on wireless communication systems is, until now, unexplored territory.

\subsection{Contributions}
In this paper, we study the impacts of capacitor mismatches in SAR ADCs on massive MU-MIMO wireless systems. 
To this end, we first propose the effective resolution (EFR), a new figure of merit (FoM) that better reflects the inherent trade-off between quantization noise and clipping artifacts than traditional metrics, such as the effective number of bit (ENOB)~\cite{Walden99}.
We then use Bussgang's decomposition~\cite{Bussgang52} to model most-significant-bit (MSB) capacitor mismatches, and we analyze the impacts of quantization, clipping, and capacitor mismatch on the EFR of a single SAR ADC. 
Finally, we present simulation results for a quantized massive MU-MIMO wireless system in order to demonstrate that capacitor mismatches in SAR ADCs can have a significant performance impact and should therefore \emph{not} be ignored.


\subsection{Notation}

Matrices are written in bold uppercase, column vectors in bold lowercase, and sets in calligraphic letters; the Frobenius norm of a matrix $\bA$ is $\frobnorm{\bA}$.
Expectation is denoted by $\Ex{\cdot}$.
We use $\phi_{\sigma_{\!X}}(\cdot)$ to refer to the probability density function (PDF) of a zero-mean Gaussian random variable $X\!\sim\! \normal(0, \sigma_X^2)$ with variance $ \sigma_X^2$.
The PDF of a standard Gaussian $\normal(0, 1)$ is $\phi(\cdot)$, and $Q(\cdot)= \int_{x}^\infty\! \phi(x) \text{d}x$ is the Q-function.
We write $\mathrm{u}(\cdot)$ for the unit step function, $\delta(\cdot)$ for the Dirac distribution, and $\mathrm{rect}_{[a,b]}(\cdot)$ for the rectangle function that is $1$ on $[a,b]$ and $0$ elsewhere.
The derivative of a function~$f$ is denoted by~$f'$.

\section{Prerequisites}

We start by analyzing the performance of a single, ideal ADC, where we only model quantization and clipping.
To this end, we first summarize Bussgang's decomposition \cite{Bussgang52}, which we use in our performance analysis. We then introduce a meaningful FoM in order to assess the performance of such an ideal ADC. 
The analysis of a SAR ADC with capacitor mismatch is provided in \fref{sec:SARmismatch}. 

\subsection{A Primer on Bussgang's Decomposition} \label{sec:buss_model}

Bussgang's decomposition models the output of a non-linear input-output transfer function (TF) $f$ applied to a zero-mean Gaussian random variable~$X$ with variance~$\sigma_{\!X}^2$ as a superposition of a linear term scaled with Bussgang gain $\beta$ and a statistically uncorrelated distortion $D$ as~\cite{Demir21}
\be \label{eq:buss_general}
f(X)=\beta X+D.
\ee
Here, $f$ will eventually model the ADC's TF.
By multiplying both sides of \fref{eq:buss_general} by $X$ and evaluating its expected value, it can be shown that the Bussgang gain $\beta$ is given by
\be\label{eq:buss_general_B}
\beta \overset{(a)}{=}\frac{\Exop\!\left[X f(X)\right]\!}{\sigma^2_{\!X}}\overset{(b)}{=}\Exop\!\left[f'(X)\right]\!.
\ee
Here, $(a)$ follows from the requirement of $D$ being uncorrelated with $X$  and $(b)$ from Stein's lemma~\cite{Stein81}. Note that $(a)$ holds for any zero-mean random variable $X$ whereas $(b)$ requires~$X$ to be Gaussian.
In order to assess the performance of a nonlinear function (e.g., the quantizer of an ADC), we are interested in characterizing the power of the distortion:\footnote{Note that the distortion power $\Exop\!\left[D^2\right]\!$ might include a bias term if $f(X)$ has nonzero mean. This bias could be removed by subtracting $\Exop\!\left[f(X)\right]\!^2$ from $\Exop\!\left[D^2\right]$, but is ignored throughout the paper for the sake of simplicity.}
\be \label{eq:buss_general_D}
\Exop\!\left[D^2\right]\!=\Exop\!\left[(f(X)-\beta X)^2\right]\!=\Exop\!\left[f^2(X)\right]\!-\beta^2\sigma_{\!X}^2.
\ee
Note that $\beta$ from \fref{eq:buss_general_B} also minimizes $\Exop\!\left[D^2\right]$.
From \fref{eq:buss_general_B} and \fref{eq:buss_general_D}, we notice that only two quantities need to be known in order to evaluate the power of the distortion~$D$, namely $\Exop\!\left[f'(X)\right]\!$ to compute the Bussgang gain $\beta$ and $\Exop\!\left[f^2(X)\right]\!$.
%

\subsection{Model for an Ideal ADC} \label{sec:buss_perf_adc}

We will now derive Bussgang's decomposition for an ideal $N$-bit uniform symmetric mid-rise quantizer with an input range of $[-1, 1]$; any value outside this range will be clipped. This ideal ADC model will be used in \fref{sec:fom} to introduce the FoMs and also serve as a baseline. 
For $x \geq 0$, we have
\begin{align} 
f_{\geq 0}(x)&=\frac{\Delta}{2}\mathrm{u}(x) + \Delta \!\!\!\! \sum_{k=1}^{2^{N-1}-1}\!\!\!\mathrm{u}(x-k\Delta), \label{eq:ideal_quant_f} \\
f'_{\geq 0}(x)&=\frac{\Delta}{2}\delta(x) + \Delta \!\!\!\! \sum_{k=1}^{2^{N-1}-1}\!\!\!\delta(x-k\Delta), \label{eq:ideal_quant_f'}
\end{align}
where $\Delta=2^{1-N}$ is a least significant bit (LSB) step width.
Note that it is sufficient to consider the case $x\geq 0$ as $f$ is an odd function; this also means that $f'$ is an even function. 
Splitting equations with the sign of $x$ is not necessary for the analysis of an ideal ADC, but will significantly simplify our derivations when considering mismatches in \fref{sec:SARmismatch}.

Using \fref{eq:ideal_quant_f'}, we can write \fref{eq:buss_general_B} as follows:
\begin{align} \label{eq:ideal_quant_B}
\beta 	& = \int_{-\infty}^{+\infty} f'(x)\phi_{\sigma_{\!X}}(x)\,\text{d}t 
	 \overset{(a)}{=} 2  \int_{0}^{+\infty} f'_{\geq 0}(x)\phi_{\sigma_{\!X}}(x)\,\text{d}t	\nonumber \\
	& \overset{(b)}{=} \frac{\Delta}{\sigma_{\!X}} \!\left(\phi(0) + 2 \sum_{k=1}^{2^{N-1}-1}\!\!\!\phi\!\left(\frac{\Delta}{\sigma_{\!X}} k\right)\!\right)\!.
\end{align}
Here, $(a)$ exploits symmetries in the functions $f'$ and $\phi_{\sigma_{\!X}}$, and $(b)$ follows from the fact that $\phi_{\sigma_{\!X}}(t)=\frac{1}{\sigma_{\!X}}\phi\big(\frac{t}{\sigma_{\!X}}\big)$.

We note that \fref{eq:ideal_quant_B} is a partial Riemann integral of $\phi_{\sigma_{\!X}}$ on the ADC range $[-1, 1]$. When taking the limit  $N\!\to\!\infty$, $\beta$ corresponds to $1-2Q(1/\sigma_{\!X})$, which is the Bussgang gain for an ADC with infinite resolution and clipping only~\cite{Bussgang52}.

With the expression for $\beta$ in \fref{eq:ideal_quant_B}, we can now follow a similar procedure to evaluate $\Exop\!\left[f^2(X)\right]\!$. We first write
\begin{align} \label{eq:ideal_quant_f2}
f^2_{\geq 0}(x) = &  \sum_{k=0}^{2^{N-1}-1}\!\! \left(\frac{\Delta}{2}+k\Delta\right)^2 \!\!\! \mathrm{rect}_{[k\Delta , (k+1)\Delta]}(x) \nonumber \\
										    &+ \!\left( 1-\frac{\Delta}{2} \right)^2 \!\!\! u(x-1),
\end{align}
from which we obtain the following result:
\begin{align} \label{eq:ideal_quant_f2_ex}
\Exop\!\left[f^2(X)\right]  =&   2 \sum_{k=1}^{2^{N-1}-1} \!\left( \left( \frac{\Delta}{2}+k\Delta \right)^2 \!\!\! \int_{k\frac{\Delta}{\sigma_{\!X}}}^{(k+1)\frac{\Delta}{\sigma_{\!X}}}\!\!\!\! \phi(t) \text{d}t \right)  \nonumber \\
												& +2  \!\left( 1-\frac{\Delta}{2} \right)^2 \!\!\! Q\!\left(\frac{1}{\sigma_{\!X}}\right)\!.
\end{align}

Now that we have an expression for the Bussgang gain $\beta$ and $\Exop\!\left[f^2(X)\right]\!$, we can insert these expressions into \fref{eq:buss_general_D} to obtain an analytical expression for the distortion power~$\Exop\!\left[D^2(X)\right]$.

\subsection{Figures of Merit} \label{sec:fom}

In order to analyze the effects of non-idealities in an ADC, it is key to utilize a meaningful FoM.
A common metric used to analyze the error introduced by a nonlinear distortion and (possibly) noise is the mean-squared error (MSE) defined as $\textit{MSE} \define \Exop\!\left[(f(X)-X)^2\right]$.
The MSE directly compares $f(X)$ with $X$, which suffers from the limitation that $f(X)$ and $X$ remain to be correlated, which is prone to  overestimating the effect of the distortion caused by nonidealities.

In order to circumvent the  limitations of the MSE, we consider the signal-to-distortion ratio (SDR) defined as 
\be \label{eq:fom_sdr}
\mathrm{\textit{SDR}}	\define	\frac{\Exop\!\left[(\beta X)^2\right]\!}{\Exop\!\left[(f(X)- \beta X)^2\right]\!} = \frac{\beta^2 \sigma_{\!X}^2}{\Exop\!\left[D^2\right]\!},
\ee   
where the denominator is the error when the scaling effect of Bussgang's decomposition in \fref{eq:buss_general} is taken into account.

Another common FoM used in the ADC literature, known as the effective number of bits (ENOB)~\cite{Walden99}, measures the effective resolution that the converter allows to reach when quantization, noise, and non-linearities are all taken into account. 
The ENOB refers to the resolution an ideal noiseless quantizer needs to have in order to reach the same signal-to-noise-and-distortion ratio (SNDR) as the ADC we want to test when a full-scale sinusoid is applied at its input.
Two important assumptions the ENOB makes is that the quantization error is independent of the input and uniformly distributed in $[-\Delta /2,+\Delta /2]$.
While the ENOB presents some practical advantages (e.g., it can be measured with a full-scale sinusoid at the input), it is not well-suited as a performance indicator for communication systems for the following two reasons. 
First, the ADC input signals are typically not sinusoids but rather Gaussian-like~\cite{Studer10}. Second, the ENOB lacks a dependency on the input amplitude (or gain), which renders it incapable of characterizing the inherent trade-off between quantization artifacts and amplitude clipping.

In order to circumvent the limitations of ENOB, we can utilize the SDR from \fref{eq:fom_sdr} to postulate a novel FoM that is related to resolution. 
Concretely, we propose the effective resolution EFR (measured in bits) defined as follows:
\begin{align} \label{eq:fom_efr}
\mathrm{\textit{EFR}}	 \define \frac{\mathrm{\textit{SDR}_\text{dB}}\!+\!10\log_{10}(2(\pi \!-\!2))}{20\log_{10}(2)}  
					\approx \frac{\mathrm{\textit{SDR}_\text{dB}}\!+\!3.59}{6.02}
\end{align}
with $\textit{SDR}_\text{dB}=10\log_{10}(\textit{SDR})$. The denominator ensures that halving the amplitude of the distortion compared to the one of the wanted signal leads to one additional bit of resolution (this principle is also used in the ENOB definition). 
The offset in the numerator ensures that a $1$-bit quantizer has an EFR of~$1\,$b.
Note that the SDR of a $1$-bit quantizer does not depend on the input amplitude (determined by $\sigma_{\!X}$), which makes the correction factor meaningful when considering different input~gains.

\begin{figure}[tp]
\centering
\includegraphics[width=0.8\columnwidth]{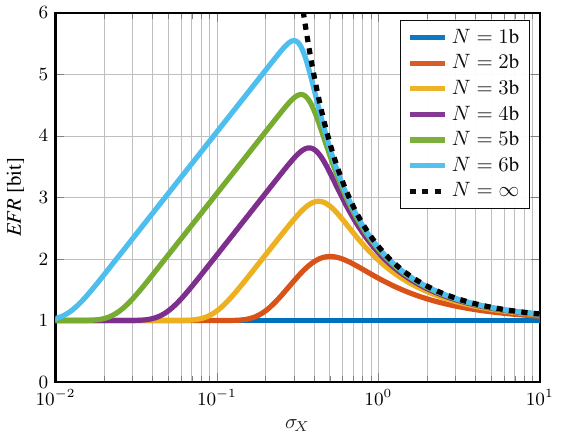}
\vspace{-0.25cm}    
\caption{Effective resolution (EFR) of an ideal $N$-bit ADC with quantization and clipping only, dependent on the input standard deviation $\sigma_{\!X}$.}
\label{fig:EFR_ideal}
\end{figure}

\subsection{EFR of an Ideal ADC} \label{sec:efr_ideal}

In order to analyze the effect of non-idealities in ADCs, we need an ideal reference. 
For this purpose, we now briefly discuss the EFR for the $N$-bit ideal ADC with input range $[-1,+1]$, where we only consider quantization and clipping as in \fref{sec:buss_perf_adc}. 
In \fref{fig:EFR_ideal}, we see that an $N$-bit quantizer reaches an EFR close to $N$ only for low resolutions. For higher resolutions, the quantization levels close to the clipping boundaries $-1$ and $+1$ cannot be fully exploited as an increase in input power would lead to more clipping artifacts. 
We also see that there exists an optimum input amplitude (determined by~$\sigma_X$ in \fref{fig:EFR_ideal}) that trades off quantization errors versus clipping artifacts---this optimum depends on the ADC's resolution.
%


\section{SAR ADC Mismatch Model} \label{sec:SARmismatch}

We now briefly present the operating principle of a successive approximation register (SAR) ADC and discuss the origin of capacitor mismatches. We then propose a mathematical model to analyze the impact of such mismatches. Finally, we present numerical results for both the model and simulations.

\subsection{Basics of SAR ADC Operation }\label{sec:sar}

\begin{figure}[tp] 
	\centering
	\includegraphics[width=0.49\textwidth]{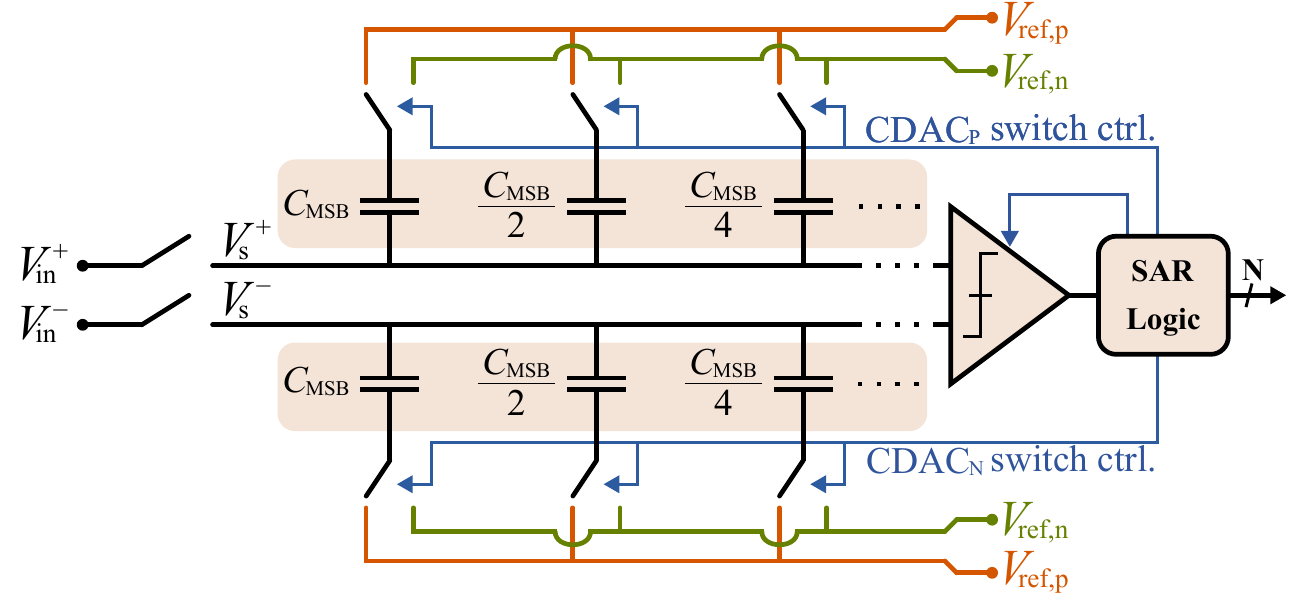}
	\vspace{-0.5cm}    
	\caption{A high-level block diagram of a SAR ADC.}
	\label{fig:SAR_blocks}
\end{figure}

A SAR ADC operates in two phases. 
In the first phase (the tracking phase), the ADC follows the input voltage and samples it at the end of this phase when the ADC sampling switches open.
In the second phase (the conversion phase), the ADC performs a binary search to sequentially determine the bits associated with the sampled voltage.
The way of implementing this binary search is charge-sharing, achieved by capacitor-array (referred to as CDAC for capacitive digital-to-analog converter) switching. 
\fref{fig:SAR_blocks} shows a high-level block diagram of a SAR ADC.
Like most modern analog circuits, SAR ADCs are implemented differentially (as opposed to single-ended implementation).
This means that the voltage we want to digitize is no longer the voltage on a single node, but the voltage difference between a positive and a negative node:
\be \label{eq:diff}
V_{\text{\text{in}},\text{d}} = V_{\text{in}}^+\! - V_{\text{in}}^-		\quad \text{and} \quad	V_{\text{s},\text{d}} = V_\text{s}^+\! - V_\text{s}^-.
\ee
The input range of the ADC is $[-\Delta V_{\text{ref}},+\Delta V_{\text{ref}}]$ with $\Delta V_{\text{ref}}=V_{\text{\text{ref}},\text{p}}-V_{\text{\text{ref}},\text{n}}$.

During the tracking phase, the sampling switches (on the left of~\fref{fig:SAR_blocks}) are closed and all capacitors are connected to $V_{\text{\text{ref}},\text{p}}$.
After the sampling switches open, the voltage is captured in the CDAC and conversion begins: $V_{\text{s},\text{d}}$ is a sampled version of~$V_{\text{\text{in}},\text{d}}$, and the comparator can convert the most significant bit (MSB).
If this first conversion yields a 1, i.e, if $V_{\text{s},\text{d}}>0$, then we need to subtract $\Delta V_{\text{ref}}/2$ from $V_{\text{s},\text{d}}$ to continue with the binary search.
To do so, we switch the MSB capacitor of~$\text{CDAC}_\text{P}$ to~$V_{\text{\text{ref}},\text{n}}$ ($\text{CDAC}_\text{P}$ is the upper capacitive array on \fref{fig:SAR_blocks}).
If the MSB was $0$, then we would do the same but on $\text{CDAC}_\text{N}$ (lower capacitive array on \fref{fig:SAR_blocks}) to add $\Delta V_{\text{ref}}/2$ to $V_{\text{s},\text{d}}$.
One can then perform the next comparison, and conditionally add or subtract $\Delta V_{\text{ref}}/4$ by switching $\text{C}_{\text{MSB}}/2$ to $V_{\text{\text{ref}},\text{n}}$ on the corresponding CDAC (depending on the comparison result). This process repeats until the required number of bits has been converted.
Note that the discussed switching scheme is  common in high-speed designs as it only requires one reference voltage ($V_{\text{\text{ref}},\text{n}}$ is typically connected to ground), but other possibilities exist.\footnote{The mismatch model presented in \fref{sec:mis_model} can still be used for other switching strategies or some other types of ADCs.
%
For example, residue amplifier mismatches of a pipelined ADC can also be modeled as capacitor mismatches.}

\subsection{Mismatches in SAR ADCs}\label{sec:sar_mis}

When a SAR ADC is manufactured, the capacitors in the CDACs do not have the exact (ideal) value they have been designed for due to process variation~\cite{Plassche94}.\footnote{Comparator offset is another typical source of mismatch in SAR ADCs, but its effect only shifts the entire TF by the amount of voltage mismatch the comparator exhibits. This source of mismatch does not cause nonlinear distortion and it is, thus, common to fix its effect with digital post-processing.}
Even if such mismatches are a random process, there will be no averaging effect: once a chip has been manufactured, its mismatches are fixed and will affect the system in a deterministic way. 

It is common to model the relative variation  $\Delta\text{C}/\text{C}$ of a capacitor's capacitance value due to mismatch by a Gaussian process with standard deviation~\cite{Pelgrom89,Marin07}
%
\be \label{eq:pelgrom}
	\sigma\!_{\Delta\text{C}/\text{C}} = \frac{A_\text{C}}{\sqrt{WL}},
\ee
where $W$ and $L$ are the physical dimensions of the capacitor, and $A_\text{C}$ is a process-dependent constant.
To minimize capacitor mismatches, unit-capacitors are typically used, where the smallest capacitor is implemented multiple times and connected in parallel to create larger capacitors. Hence, doubling a capacitor's size is equivalent to doubling its area $W\! L$, therefore dividing its relative variations by $\sqrt{2}$.

However, during the binary search performed by a SAR ADC, a capacitor switched from $V_{\text{\text{ref}},\text{p}}$ to $V_{\text{\text{ref}},\text{n}}$ induces a voltage change to $V_{\text{s},\text{d}}$ proportional to its size.
The largest capacitor $\text{C}_{\text{MSB}}$ has $\sqrt{2}$ times less variation than its subsequent capacitor $\text{C}_{\text{MSB}}/2$ on the CDAC, but induces a voltage change twice as large.
Therefore, $\text{C}_{\text{MSB}}$ mismatch induces a voltage error $\sqrt{2}$ times larger than its subsequent capacitor.
The same reasoning holds every time the capacitor size scales down by a factor of two in the CDAC. Due to this behavior, MSB capacitor mismatch has the largest influence on the conversion result.

\begin{figure}[tp]
	\centering
	\includegraphics[width=0.8\columnwidth]{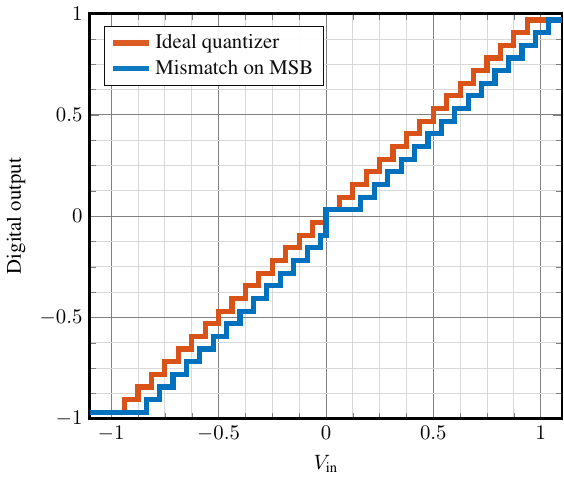}
	\vspace{-0.25cm}
	\caption{Input-output function of a $5$b-SAR ADC with $25\%$ MSB mismatch.}
	\label{fig:SAR_inout}
\end{figure}

\fref{fig:SAR_inout} shows the input-output TF $f$ of a 5b-SAR ADC with $25\%$ mismatch on its MSBs. Note that this is a large mismatch value used for illustration purposes only.
The positive array $\text{CDAC}_\text{P}$ has a $+25\%$ increase on its $\text{C}_{\text{MSB}}$, while $\text{CDAC}_\text{N}$ has $-25\%$.
Before discussing the shape of the TF $f$, we make the observation that the MSB mismatch on the $\text{CDAC}_\text{P}$ and $\text{CDAC}_\text{N}$ can only influence the conversion outcome for positive and negative input voltages respectively: this is due to the fact that the MSB is the sign of the input.

When the mismatch is positive, the ADC over-corrects during the first quadrant-shift of the binary-search.
This leads to input values close to zero being compressed to the closest-to-zero output, creating a saddle-point (see the $x \geq 0$ side in \fref{fig:SAR_inout}). It also extends the range of the ADC.
%
%
When the mismatch is negative, the ADC under-corrects the first quadrant-shift. Outputs close to zero are then unreachable as shown on the $x < 0$ side in \fref{fig:SAR_inout}.
Inputs close to the limits of the ADC range will also be clipped as the under-correction makes them larger than they should after the first shift.
Overall, digital outputs close to zero are unreachable and the ADC's range is reduced.
We also see in \fref{fig:SAR_inout} that mismatches can lead to a fixed shift of the TF---to the right in the case illustrated here.

\subsection{Model for SAR ADC with MSB-Mismatch}\label{sec:mis_model}

\begin{figure}[tp]
	\centering
	\includegraphics[width=0.8\columnwidth]{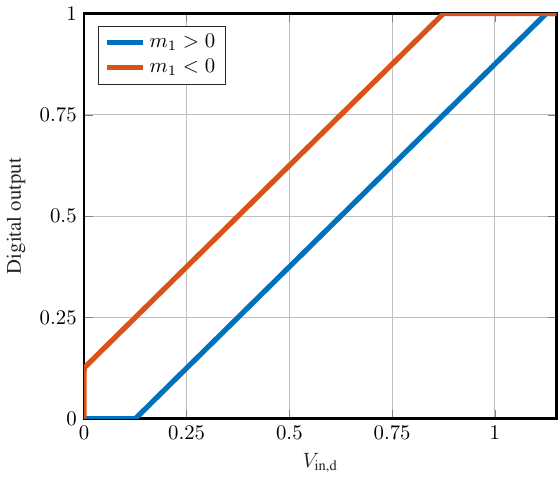}
	\vspace{-0.25cm}
	\caption{Transfer functions depending on the sign for MSB mismatch modeling.}
	\label{fig:mis_function}
\end{figure}

As explained in \fref{sec:sar_mis}, MSB mismatch has the strongest impact on the final conversion result. To disentangle the effect of the MSB mismatch from the effect of quantization itself, we will model the ADC TF as a piecewise linear function that includes the effect of clipping and the displacement from the ideal TF due to MSB-only mismatch.
We will denote $m_1$ the TF shift induced by mismatch of the MSB on the positive CDAC and $m_2$ the one on the negative CDAC
\be \label{eq:def_m}
m_1 = \frac{1}{2} \frac{\Delta \text{C}_{\textit{MSB},+}}{\text{C}_{\textit{MSB},+}}	\quad \text{and} \quad m_2 = \frac{1}{2} \frac{\Delta \text{C}_{\textit{MSB},-}}{\text{C}_{\textit{MSB},-}},
\ee
where the $1/2$ factor comes from the first quadrant shift amplitude of $\Delta V_{\text{ref}}/2$.
For $x\ge 0$, the transfer function $f$ for both possible signs of $m_1$ are
\begin{align} \label{eq:sar_mis_f}
\begin{split} 
	&f_{m_1 \ge 0}(x)=(x\!-\!m_1) \rect_{[m_1,1+m_1]}(x) \!+\! \mathrm{u}(x\!-\!1\!-\!m_1) \\
	&f_{m_1 < 0}(x)=(x\!-\!m_1) \rect_{[0,1+m_1]}(x) \!+\! \mathrm{u}(x\!-\!1\!-\!m_1),
\end{split}
\end{align}
where the resulting functions are shown in \fref{fig:mis_function} for a $25\%$ capacitor mismatch, i.e., for $\!\left|m_1\right|\!=0.125$.

On the $x<0$ side, we can write the same equations by replacing $m_1$ by $m_2$ and mapping $f(x)$ to $-f(-x)$.
However, there is no need to write this explicitly as we can simply re-assemble the different component together at the end of the calculation as done for \fref{eq:ideal_quant_B} and \fref{eq:ideal_quant_f2_ex}.
Now that we have the mismatched TF~$f$, we take its derivative:
\begin{align} \label{eq:sar_mis_fprime}
\begin{split}
	f'_{m_1 \ge 0}(x)&=\rect_{[m_1,1+m_1]}(x) \\
	f'_{m_1 < 0}(x)&=\rect_{[0,1+m_1]}(x)-m_1\delta(x).
\end{split}
\end{align}

After integration, we can calculate the Bussgang gain
\be \label{eq:betaOMGcrazymanycases}
	\beta=
	\begin{cases}
		Q(\frac{m_1}{\sigma_{\!X}})-Q(\frac{1+m_1}{\sigma_{\!X}}) \\
		\quad+Q(\frac{m_2}{\sigma_{\!X}})-Q(\frac{1+m_2}{\sigma_{\!X}}), & m_1\ge 0, m_2\ge 0 \medskip\\
		
		Q(\frac{m_1}{\sigma_{\!X}})-Q(\frac{1+m_1}{\sigma_{\!X}}) \\
		\quad+\frac{1}{2}\!-\!Q(\frac{1+m_2}{\sigma_{\!X}})-
		\frac{m_2}{\sigma_{\!X}}\phi(0), \!\!\!\!\! & m_1\ge 0, m_2< 0 \medskip\\
		
		\frac{1}{2}-Q(\frac{1+m_1}{\sigma_{\!X}})-\frac{m_1}{\sigma_{\!X}}\phi(0)\\
		\quad+Q(\frac{m_2}{\sigma_{\!X}})-Q(\frac{1+m_2}{\sigma_{\!X}}), & m_1<0, m_2\ge 0 \medskip\\
		
		1-Q(\frac{1+m_1}{\sigma_{\!X}})-Q(\frac{1+m_2}{\sigma_{\!X}}) \\
		\quad-\frac{m_1+m_2}{\sigma_{\!X}} \phi(0), & m_1<0, m_2<0.
	\end{cases}
\ee

To compute the SDR, we also need $\Ex{f^2(X)}$. We have
\begin{align} \label{eq:sar_mis_f2}
\begin{split}
	\!f^2_{m_1 \ge 0}(x) &\!=\!(x\!-\!m_1)^2  \rect_{[m_1,1+m_1]}(x) \!+\! \mathrm{u}(x\!-\!1\!-\!m_1) \\
	\!f^2_{m_1 < 0}(x) &\!=\!(x\!-\!m_1)^2 	 \rect_{[0,1+m_1]}(x) \!+\! \mathrm{u}(x\!-\!1\!-\!m_1)
\end{split}
\end{align}
and can calculate $\Exop\!\left[f^2(X)\right]$ by integrating only the $x \geq 0$ side and add together the separate $m_1$ and $m_2$ component to obtain the final result.
Let us consider the following example with both mismatches being negative and $F(x,m)=f_m^2(x)$ to put the emphasis on the $m$ dependency: 
\begin{align} \label{eq:sar_mis_sym}
	\opE\!\left[f^2(X)|m_1,m_2<0\right]\! = \int_{-\infty}^{0}\!\!\!\!F_{m<0}(-t,m_2)\phi_{\sigma_{\!X}}(t)\,\text{d}t&\nonumber\\
	+\int_{0}^{+\infty}\!\!\!\!F_{m<0}(t,m_1)\phi_{\sigma_{\!X}}(t)\,\text{d}t&\nonumber\\
	=  \int_{0}^{+\infty}\!\!\!\!\left(F_{m<0}(t,m_2)+F_{m<0}(t,m_1)\right)\!\phi_{\sigma_{\!X}}(t)\,\text{d}t.&
\end{align}

Hence, we only integrate for positive values of $x$:
\begin{align} \label{eq:sar_mis_f2_integ1}	
	\Exop\!\left[f^2_{m_1 \ge 0}(X)\right]\! =(\sigma_{\!X}^2+m_1^2) \!\left(\!Q\!\left(\frac{m_1}{\sigma_{\!X}}\right)\!-Q\!\left(\frac{1\!+\!m_1}{\sigma_{\!X}}\right)\!\right)&  \nonumber\\
	+Q\!\left(\frac{1\!+\!m_1}{\sigma_{\!X}}\right)\! + \sigma_{\!X} \!\left(\!(m_1-1) \phi\!\left(\frac{1\!+\!m_1}{\sigma_{\!X}}\right)\!-m_1 \phi\!\left(\frac{m_1}{\sigma_{\!X}}\right)\!\right)&
\end{align}

and
\begin{align} \label{eq:sar_mis_f2_integ2}
	& \Exop\!\left[f^2_{m_1 < 0}(X)\right]\! =\frac{\sigma_{\!X}^2+m_1^2}{2} + Q\!\left(\frac{1\!+\!m_1}{\sigma_{\!X}}\right)\! (1-\sigma_{\!X}^2-m_1^2) \nonumber\\
	& \qquad +\sigma_{\!X} \!\left((m_1\!-\!1)\phi\!\left(\frac{1\!+\!m_1}{\sigma_{\!X}}\right)\!-2m_1 \phi(0)\right).
\end{align}
With those last results, we obtain an expression for $\Exop\!\left[f^2(X)\right]\!$ by adding the $m_1$ and $m_2$ component as shown in~\fref{eq:sar_mis_sym}. Finally, the distortion power $\Exop\!\left[D^2\right]\!$ is obtained by inserting~\fref{eq:sar_mis_f2_integ2} together with the Bussgang gain from \fref{eq:betaOMGcrazymanycases} into~\fref{eq:buss_general_D}. 

\subsection{EFR of a Mismatched SAR ADC} \label{sec:adc_numerical}

\begin{figure*}[tp]
\centering
	\subfigure[Fixed $m=0.125$.]{\includegraphics[width=0.3\linewidth]{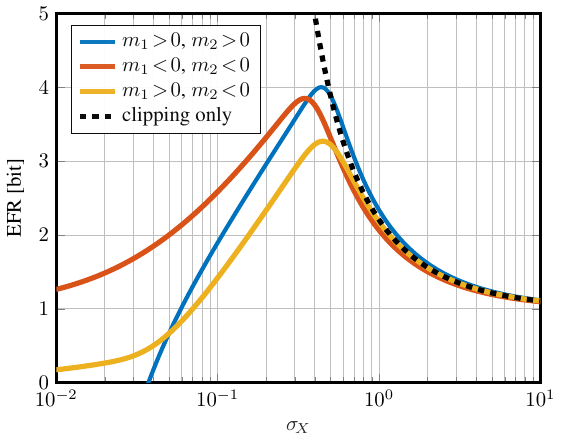}\label{fig:model_fix_m}}
\hfill
	\subfigure[Peak achievable EFR as a function of $m$.]{\raisebox{-0.06cm}{\includegraphics[width=0.33\linewidth]{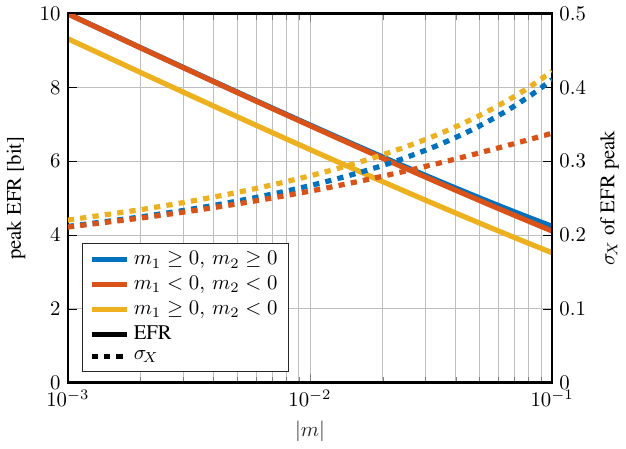}\label{fig:model_var_m}}}
\hfill
	\subfigure[$10\%$-quantile EFR of a $4$\,b mismatched ADC.]{\includegraphics[width=0.284\textwidth]{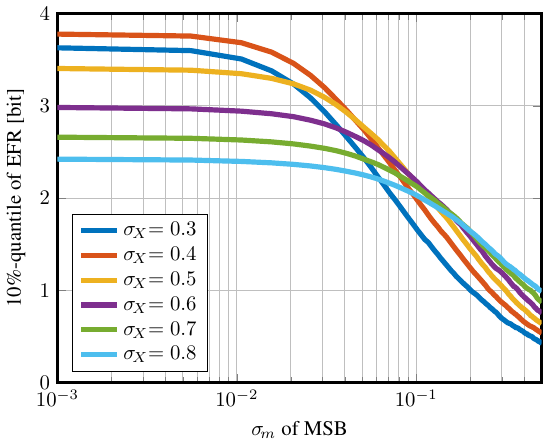}\label{fig:Sim_decile}}
\vspace{-0.15cm}  
\caption{EFR numerical results from the model with clipping and $|m_1|=|m_2|=m$ of MSB mismatch (a) and (b); simulation results for a SAR-ADC with mismatches on \emph{all} CDAC-capacitors (c).}
\end{figure*}

With our model from \fref{sec:mis_model}, we can directly compute the EFR for a given mismatch on the MSB for any input gain~$\sigma_{\!X}$.
\fref{fig:model_fix_m} shows the EFR we can reach without quantization or clipping, and for a fixed mismatch on both MSB capacitors of $|m_{1,2}|=0.125$ as described in \fref{sec:mis_model}.
Note that the mismatch value corresponds to the size of an LSB for a $4$-bit data converter.
In general, it is making sense to compare a mismatch value to the size of an LSB step width from the quantizer we consider.
This consideration is typical for analog designers while implementing a SAR ADC: one aims at limiting variations to a fraction of an LSB, as it avoids missing output codes and ensures monotonicity in the TF of the data~converter~\cite{Plassche94}.

\fref{fig:model_var_m} shows the maximum achievable EFR as a function of~$|m|$. We also show at which input gain $\sigma_{\!X}$ this peak EFR is achieved. We first observe that the optimal gain $\sigma_{\!X}$ is increasing with $|m|$.
This is due to the trade-off between the non-linearity around zero and clipping artifacts: stronger mismatch makes it desirable to allow for  larger clipping artifacts to avoid the jump (or saddle point) around zero.
Examining \fref{fig:model_var_m} closely also reveals that the slope of peak EFR is of $-1\,$bit/octave.
This corroborates that comparing $|m|$ with the size of an LSB step is a meaningful procedure: halving the MSB mismatch leads to one additional bit of resolution, which is equivalent to halving the size of the LSB step width.

A $4$-bit mid-rise SAR ADC following the model discussed in \fref{sec:buss_perf_adc} with random mismatch on \emph{all} its CDAC-capacitors has been simulated.
The standard deviation $\sigma_{\!m}$ of the random mismatch affecting the MSB is defined as in \fref{eq:def_m}.
As explained in \fref{sec:sar}, the standard deviation of capacitor variations is divided by $\sqrt{2}$ every time the capacitance value is halved along the CDAC.
Results are shown in \fref{fig:Sim_decile}, where we plot the $10\%$-quantile of the achieved EFR over CDAC-capacitors mismatch realizations.
As mismatches are fixed once a chip is fabricated, one needs to consider a worst-case metric that will determine the production yield.
We use the $10\%$-quantile which leads to $90\%$ of fabricated ADCs exhibiting an EFR greater or equal than what is shown in \fref{fig:Sim_decile}.

With the $10\%$-quantile measure, we notice that capacitor mismatch starts to significantly impact the EFR when $\sigma_{\!m}$ is of the order of a tenth of an LSB ($\Delta=0.125$ in our $4$-bit case). Similar results were obtained for all other resolutions in the simulated range ($2$\,b to $6$\,b).
As for the case of the MSB-mismatch-only model, we notice that the gain $\sigma_{\!X}$ leading to the highest EFR increases with $\sigma_{\!m}$, but, due to quantization, with a bounded maximum EFR.
The maximum value reached at low mismatch standard deviations~$\sigma_{\!m}$ corresponds to what was shown theoretically in \fref{sec:efr_ideal} (cf.~\fref{fig:EFR_ideal}).
In summary, our simulation results confirm the intuition gained from the proposed mismatch model, and show the impact of using a worst-case measure like the $10\%$-quantile. 
%


\section{Impact of SAR ADC Mismatch on Quantized Massive MU-MIMO Systems}\label{sec:mimo_sim}

As shown in \fref{sec:SARmismatch}, capacitor mismatches can have a considerable impact on the effective resolution of a single SAR ADC. 
We now show that the same holds true for a quantized massive MU-MIMO wireless system, meaning that SAR ADC mismatches should \emph{not} be ignored. 
To this end, we first introduce the system model and, then, show simulation results for the case of 4-bit ADCs with mismatches. 

\subsection{System Model}

We consider a mmWave massive MU-MIMO uplink scenario in which $U=16$ single-antenna user equipments (UEs) transmit data to a $B=64$ antenna basestation (BS) with a uniform linear array.
We assume that the UEs utilize power control to stay within $\pm 3$\,dB from the average power $\Es$.  
We use a QuaDRiGa mmMagic channel model~\cite{jaeckel2019quadriga} with $1^\circ$ minimum angular separation between the UEs.
We assume that the BS has perfect channel state information and uses a linear minimum mean-square error (LMMSE) detector.
We consider the following discrete-time frequency-flat system model:
\be \label{eq:mimo_inout}
	\bmy = f(\bmz) \quad \text{with} \quad \bmz= \bH \bms + \bmn .
\ee
Here, $\bmy\in\setA^B$ is the BS receive matrix with output code alphabet $\setA$ after the ADCs transfer function $f$, which  includes quantization, mismatch, and clipping applied element-wise to the ideal receive vector $\bmz\in\complexset^B$, $\bH\in\complexset^{B\times U}$ is the effective channel matrix including power control, $\bms\in\setS^U$ contains the transmitted 16-QAM data symbols, and $\bmn\in\complexset^B$ models thermal noise and is assumed to contain i.i.d.\ circularly-symmetric complex Gaussian entries with variance $\No$. 
We define the average receive signal-to-noise ratio (SNR) as
\be	\label{eq:snr_def}
	\textit{SNR} \define \frac{\frobnorm{\bH}^2E_s}{BN_0}.
\ee

Note that the SAR ADC TF $f$ in \fref{eq:mimo_inout} is applied entry-wise to the real and imaginary part of the ideal input vector $\bmz$, which means that for $B$ BS antennas, we model an array of $2B$ SAR ADCs. 
As discussed in \fref{sec:adc_numerical}, we define $\sigma_{\! m}$ as the standard deviation of the Gaussian distribution modeling the effect of the MSB mismatch as defined in \fref{eq:def_m}.

\begin{figure*}[tp]
	\subfigure[Mismatch in all CDAC-capacitors]{\includegraphics[width=0.28\linewidth]{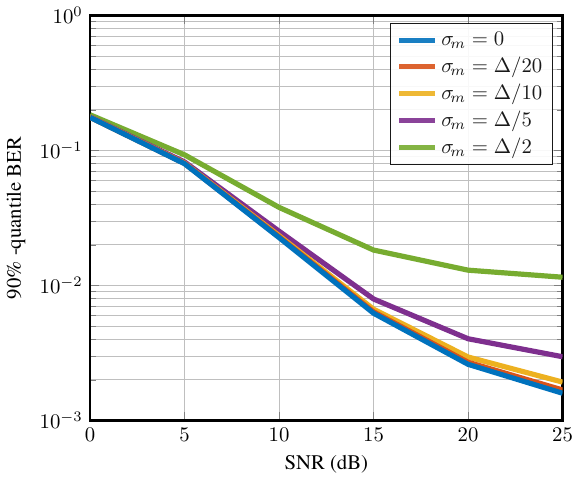}\label{fig:mimo_all}}
\hfill
	\subfigure[Mismatch on MSBs only]{\includegraphics[width=0.28\linewidth]{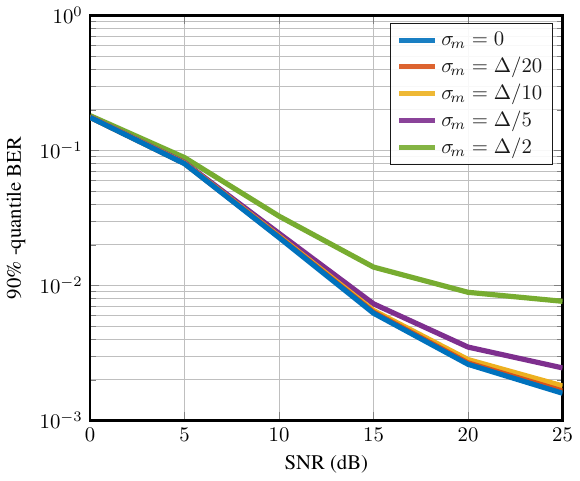}\label{fig:mimo_msb}}
\hfill
	\subfigure[CDF at $15$\,dB SNR with mismatch in all CDAC-capacitors.]{\includegraphics[width=0.28\linewidth]{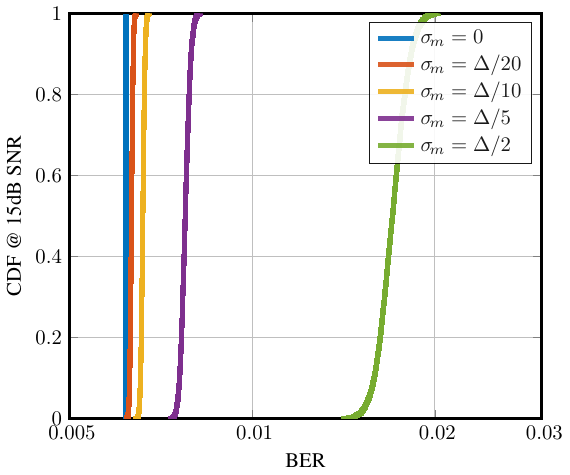}\label{fig:mimo_cdf}}
\vspace{-0.15cm}    
\caption{Massive MU-MIMO simulation with 4b SAR ADC showing the 90\%-quantile of the achieved uncoded BER.}
\end{figure*}

\subsection{Simulation Results}\label{sec:sim_results}

We now show simulation results for the case of $4$-bit SAR-ADCs. Different values for $\sigma_{m}$ have been simulated, with values being different fractions of an LSB-step $\Delta=0.125$. We use the uncoded bit error rate (BER) as our performance metric.
It is important to note that we report the 10\% worst-case uncoded BER,  the same way as in \fref{fig:Sim_decile}, because mismatch does not have any averaging property once a chip is fabricated. Since a lower uncoded BER is better than a higher one, this 10\% worst case corresponds to the $90\%$-quantile.

In \fref{fig:mimo_all}, we see the $90\%$-quantile BER when mismatch is present on all CDAC capacitors in all SAR ADCs. 
We see a significant influence of SAR ADC mismatch: A standard deviation $\sigma_{m}$ of $\Delta/2$ still leads to an uncoded BER floor of about $1\%$, which may be unacceptable for real-world wireless communication systems.

In \fref{fig:mimo_msb}, we observe similar behavior from the same simulation setup but where we only consider mismatches in the MSB capacitors. 
When comparing \fref{fig:mimo_all} to \fref{fig:mimo_msb}, we see that considering only MSB mismatch well-approximates a model with mismatches in all CDAC capacitors; this confirms the fact that only modeling the MSB mismatch, as in \fref{sec:mis_model}, is a sensible approximation. 
We also notice that, for larger mismatches, the BER difference between the MSB-mismatch-only and the all-capacitor-mismatch models increases. Such large mismatches, however, would anyway be avoided in practical systems and are therefore less relevant. 

In \fref{fig:mimo_cdf}, we show the cumulative distribution function (CDF) of the uncoded BER at an SNR of $15$\,dB. 
Throughout this work, we chose to use $\sigma_{ m}$ and a $10\%$ worst-case BER performance as our performance metric. 
Once a target BER specification has been set, all of the fabricated ADCs that are not meeting the specifications have to be discarded. The production yield of an ADC design is therefore defined by the proportion of designs not meeting the target specifications, no matter how good the rest of the designs happen to be.


\section{Conclusions}

We have analyzed the impact of capacitor mismatches caused by semiconductor process variations in widely-used successive approximation register (SAR) analog-to-digital converters (ADCs). 
We have developed an analytical model based on Bussgang's decomposition for the most-significant bit (MSB) mismatch of a SAR ADC, and we have analyzed the effects of quantization, clipping, and mismatches on a single ADC with a new figure of merit: the effective resolution (EFR).
Our EFR results for a single SAR ADC reveal trade-offs between quantization errors, mismatch distortion, and clipping artifacts that all depend on the ADC's input gain.
Finally, we have demonstrated the impact of SAR ADC mismatch on the bit error rate (BER) performance of a quantized massive MU-MIMO communication system. Our simulation results reveal that SAR ADC mismatches should \emph{not} be ignored as they significantly affect the BER, even for low-resolution ADCs.

In practice, ADC designers typically must adhere to stringent specifications in order to limit the impact of process variations on ADC performance. Such strict constraints, however, are against the goal of utilizing simple, low-cost, and energy-efficient analog front-end designs in  massive MU-MIMO systems with potentially hundreds of ADCs. 
Our work reveals that one can take capacitor mismatch into account when designing systems with SAR ADC arrays, which has the potential of extracting relaxed specifications---this paves the way for more efficient ADC array architectures.

\balance

\bibliographystyle{IEEEtran}
\bibliography{bib/VIPabbrv,bib/confs-jrnls,bib/publishers,bib/VIP_190331,bib/bibliography}

\balance

\end{document}